\RequirePackage{fix-cm}
\documentclass[pdftex,twocolumn,epjc3]{svjour3}  
\smartqed  % flush right qed marks, e.g. at end of proof
\RequirePackage[T1]{fontenc}
\RequirePackage{graphicx}
 \RequirePackage{mathptmx}      % use Times fonts if available on your TeX system
 \RequirePackage{flushend}
\RequirePackage{latexsym}
\RequirePackage{multirow}
\RequirePackage[numbers,sort&compress]{natbib}

\RequirePackage{siunitx}

\RequirePackage[colorlinks,citecolor=blue,urlcolor=blue,linkcolor=blue]{hyperref}

\RequirePackage{lineno}
\RequirePackage{amsmath}
\RequirePackage{etoolbox} %% <- for \cspreto, \csappto
\RequirePackage{microtype}

\newcommand*\linenomathpatch[1]{%
  \cspreto{#1}{\linenomath}%
  \cspreto{#1*}{\linenomath}%
  \csappto{end#1}{\endlinenomath}%
  \csappto{end#1*}{\endlinenomath}%
}

\linenomathpatch{equation}
\linenomathpatch{gather}
\linenomathpatch{multline}
\linenomathpatch{align}
\linenomathpatch{alignat}
\linenomathpatch{flalign}

\journalname{Eur. Phys. J. C}
\begin{document}

\title{Precision Measurement of the Specific Activity of $^{39}$Ar in Atmospheric Argon with the DEAP-3600 Detector
}
\subtitle{}%Do you have a subtitle?\\ If so, write it here}


\author{
        P.~Adhikari~\thanksref{Carleton}
        \and
        R.~Ajaj~\thanksref{Carleton, Mcdonaldinst}
        \and
        M.~Alp\'{i}zar-Venegas~\thanksref{UNAM}
        \and
        P.-A.~Amaudruz~\thanksref{Triumf}
        \and
        J.~Anstey~\thanksref{Carleton, Mcdonaldinst}
        \and
        G.R.~Araujo~\thanksref{TUM}
        \and
        D.J.~Auty~\thanksref{Alberta}
        \and
        M.~Baldwin~\thanksref{RAL}
        \and
        M.~Batygov~\thanksref{Laurentian}
        \and
        B.~Beltran~\thanksref{Alberta}
% ...
        (DEAP Collaboration)\thanksref{email}
}


\thankstext{deceased}{Deceased.}
\thankstext{email}{deap-papers@snolab.ca}
\thankstext{PGiampa}{Currently at SNOLAB, Lively, Ontario, P3Y 1M3, Canada}
\thankstext{TRPollman}{Currently at Nikhef and the University of Amsterdam, Science Park, 1098XG Amsterdam, Netherlands}


\institute{
    Department  of  Physics,  University  of  Alberta,  Edmonton,  Alberta,  T6G  2R3,  Canada \label{Alberta}
% ...
\and
    Department  of  Physics,  Carleton  University,  Ottawa,  Ontario,  K1S  5B6, Canada \label{Carleton}
% ...
\and
    School of Natural Sciences, Laurentian University, Sudbury, Ontario, P3E 2C6, Canada \label{Laurentian}
% ...
\and
    Instituto de F\'{i}sica, Universidad Nacional Aut\'{o}noma de M\'{e}xico, A. P. 20-364, Ciudad de M\'{e}xico 01000, Mexico \label{UNAM}
% ...
\and
    Rutherford Appleton Laboratory, Harwell Oxford, Didcot OX11 0QX, United Kingdom \label{RAL}
% ...
\and
    TRIUMF, Vancouver, British Columbia, V6T 2A3, Canada \label{Triumf}
    \and
    Department of Physics, Technische Universit\"{a}t M\"{u}nchen, 80333 Munich, Germany \label{TUM}
    \and
    Arthur B. McDonald Canadian Astroparticle Physics Research Institute, Queen's University, Kingston, ON, K7L 3N6, Canada \label{Mcdonaldinst}
}

\date{Received: date / Accepted: date}

\maketitle

\begin{abstract}
  The specific activity of the $\beta$ decay of $^{39}$Ar in atmospheric argon is measured using the DEAP-3600 detector. DEAP-3600, located 2 km underground at SNOLAB, uses a total of (3269 $\pm$ 24) kg of liquid argon distilled from the atmosphere to search for dark matter.  This detector is well-suited to measure the decay of $^{39}$Ar owing to its very low background levels.  This is achieved in two ways: it uses low background construction materials; and it uses pulse-shape discrimination to differentiate between nuclear recoils and electron recoils. With 167~live-days of data, the measured specific activity at the time of atmospheric extraction is (0.964 $\pm$ 0.001$_{\rm stat}$ $\pm$ 0.024$_{\rm sys}$) Bq/kg$_{\rm atmAr}$, which is consistent with results from other experiments. A cross-check analysis using different event selection criteria and a different statistical method confirms the result.
  \keywords{Atmospheric argon \and $\beta$ decay \and Specific activity \and $^{39}$Ar \and DEAP-3600}
\end{abstract}

\section{Introduction}

Argon is used as a target material in a variety of existing and future particle detectors \cite{fiorillo, atlasECal, darkside50, DS-20k, large, microboone, protodune, Abi_2020, detpaper}. 
Commercially available argon is obtained by distillation from the Earth's atmosphere where it has a natural abundance of about 0.93\% \cite{allensAQ}. 
While atmospheric argon primarily consists of the stable isotope $^{40}$Ar, trace amounts of cosmogenically created, radioactive $^{39}$Ar are also present and represent a background in low-threshold detectors. 
The isotope $^{39}$Ar decays via unique first-forbidden $\beta$ decay with a half-life of \mbox{$T_{1/2} = (269 \pm 9)$}~years and a Q-value of \mbox{(565 $\pm$ 5)}~keV \cite{argon39meas, arhalflife, arqvalue}. 

While the production of $^{39}$Ar in the atmosphere is in equilibrium, measurements of ice cores and tree rings by Gu et al. \cite{GUArRatios} show the $^{39}$Ar/Ar ratio has varied by as much as 17\% in the past 2500 years.
Recent measurements of the specific activity of $^{39}$Ar in atmospheric argon, $S_{\rm Ar39}$, were realized by the WARP collaboration with a result of 
$S_{\rm Ar39} = (1.01 \pm 0.02_{\rm stat} \pm 0.08_{\rm sys})$\,Bq/kg$_{\rm atmAr}$ \cite{warp}
and by the ArDM collaboration with $S_{\rm Ar39} = (0.95 \pm 0.05)$\,Bq/kg$_{\rm atmAr}$ \cite{ardm}.

This paper describes the measurement of the $^{39}$Ar specific activity using the DEAP-3600 detector~\cite{detpaper}, located 2\,km underground in Creighton Mine at SNOLAB in Sudbury, Ontario, Canada. DEAP-3600 is a dark matter experiment with a liquid argon (LAr) target which achieves low-background levels due to both its use of low-background construction materials and implementation of pulse-shape discrimination (PSD).  The PSD technique is able to differentiate between nuclear recoils and electron recoils; it achieves an expected leakage of electron recoil events into the dark matter search region of interest of fewer than 1 event per year of data. 
The large mass of atmospheric argon and the very low background levels achieved with this experiment \cite{1styearpaper,erb} enable the precision specific activity measurement presented here. The specific activity is calculated by estimating the total number of $^{39}$Ar decays $N$ within a certain live-time $T_{\rm live}$ as shown in Eq. \ref{eq:specact}.
\begin{equation}
    S_{\rm Ar39} = \frac{N}{T_{\rm live} \cdot m_{LAr}},
    \label{eq:specact}
\end{equation}
where $m_{LAr}$ is the mass of LAr in the detector. 

A brief description of DEAP-3600 is provided in Section~\ref{sec:det}. A dedicated estimate of the LAr mass in DEAP-3600 is presented in Section \ref{sec:mass}. 
The dataset and livetime calculation, as well as the event selection are described in Section \ref{sec:datasel}. 
The measurement of $N$ is presented in Section~\ref{sec:method}, alongside details on the $S_{\rm Ar39}$ calculation, the systematic uncertainties, and the results.
Section~\ref{sec:crosscheck} briefly describes an updated version of the measurement from Ref.~\cite{mattthesis} which appears here as a cross-check.
Concluding remarks are given in Section \ref{sec:conclusion}. 

\section{The Detector and Data Acquisition System}\label{sec:det}

The DEAP-3600 experiment operated an ultra-pure LAr target of over 3 tonnes held in a spherical acrylic vessel (AV) with 85\,cm radius from November 2016 to April 2020. The atmospheric argon for the LAr target was procured from Air Liquide.

Connected to the top of the AV is an acrylic neck surrounding a liquid nitrogen (LN$_2$) filled stainless steel cooling coil which condenses the gaseous argon (GAr) contained in the top of the AV. The AV was partially filled with the GAr/LAr interface approximately 55\,cm above the equator. 
After filling with LAr the detector was sealed and this volume of argon remained within the AV for the duration of data taking.  Cooling of the LAr was achieved by continuous circulation of LN$_2$ within the cooling coil.

\begin{figure}
\centering
\includegraphics[width=\columnwidth]{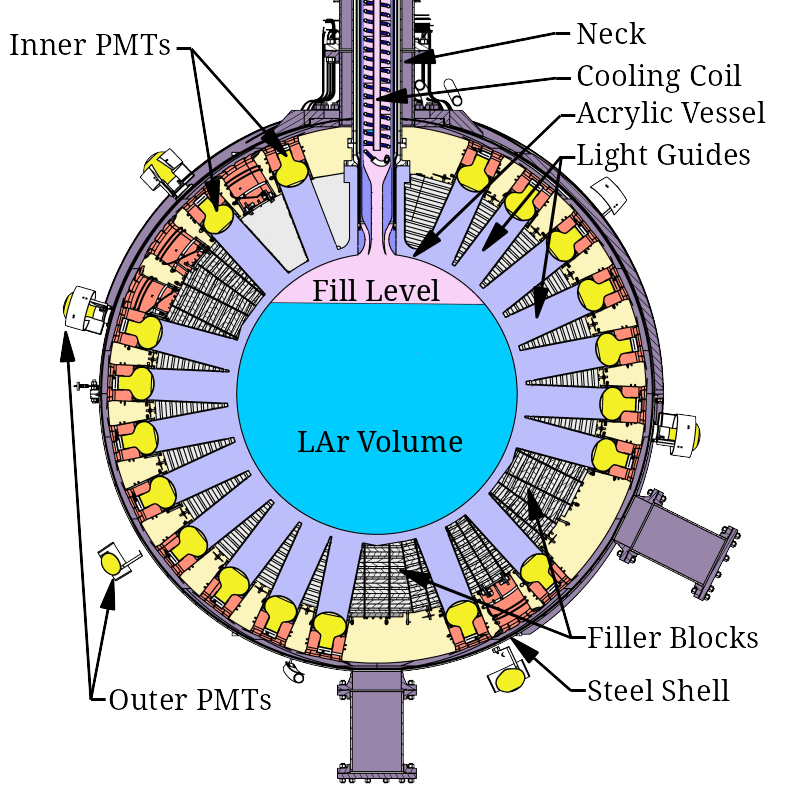}
 \caption{A cross-section of the DEAP-3600 detector components located inside the water Cherenkov muon veto detector (not shown).}
 \label{fig:detector}
\end{figure}

Coated on the inner surface of the AV is a layer of tetra-phenyl butadiene (TPB).  The TPB wavelength-shifts the 128~nm ultraviolet (UV) scintillation light from the LAr target into the visible spectrum with a peak at 420~nm \cite{tpb_2013}. This light is detected by 255 photomultiplier tubes (PMTs) which point inward and are optically coupled to the AV by acrylic light guides. The PMTs are distributed in rings around the AV, with the PMTs in each ring having the same vertical position. The AV and PMTs are enclosed in a stainless steel shell which is continuously flushed with radon-scrubbed nitrogen gas. Installed on the outer surface of the shell are 48 PMTs which point outward and, combined with the water held within a cylindrical tank surrounding the shell, act as a muon veto system.  This muon veto system detects Cherenkov light produced by muon interactions within the water. A schematic of the detector is shown in Figure \ref{fig:detector}.

Temperature sensors are placed around the AV at 85 locations along filler blocks which are mounted in the spaces between the PMTs.  The sensors are spread around the AV and placed at distances of 0.9~m, 1.1~m, or 1.3~m from the centre of the AV.  These sensors, along with the temperature and pressure within the LN$_2$ cooling system, are monitored and logged using a slow control system.

Within the data acquisition system (DAQ) each PMT is connected to a channel on a custom-built signal conditioning board (SCB) which achieves the high voltage decoupling and shapes the signals. The SCB outputs are transmitted to high-gain (CAEN V1720) and low-gain (CAEN V1740) waveform digitizers. These digitizers convert a continuous analogue signal to a discrete digital signal using analogue-to-digital converters (ADCs).
The summed input from each SCB is also passed to a digitizer and trigger module (DTM) which resolves the trigger criteria based on two rolling charge integrals: a narrow integral $Q_n$ over a 177\,ns window and a wide integral $Q_w$ over a 3.1\,$\mu$s window. The promptness of the signal is computed by the $Q_n$/$Q_w$ fraction.
Five trigger regions are defined based on these three variables. A prescaling factor of 100 is applied to events in the energy range of $Q_n \approx [50, 565]$\,keV$_{\rm ee}$ in the low $Q_n$/$Q_w$ region. This prescaling predominantly affects $^{39}$Ar decays and reduces the available statistics by storing only the observed PMT waveforms for precisely 1 out of every 100 events. In a 24 hour period, roughly $2.7\times 10^6$ $^{39}$Ar events remain after the prescaling. The timestamp of every event, included those which are prescaled, is recorded in the data.

The DTM makes the decision to trigger based on the summed value of $Q_n$ from all 255 PMTs and sends a trigger signal to the digitizers if this value passes a threshold of 19~PE. Each digitizer channel records PMT waveforms for $16$\,$\mu$s upon receiving a trigger signal, including a pre-trigger window of $2.4$\,$\mu$s. The data acquisition system is operated by \texttt{MIDAS}~\cite{midas} and the data are analyzed with \texttt{RAT}~\cite{rat}, a software framework built on \textsc{Geant4}~\cite{geant4} and \texttt{ROOT}~\cite{root}. The observed charge in each PMT is integrated over a window of [-28, 10000] ns relative to the event time. This charge is divided by the single photoelectron (PE) charge for each PMT measured through independent calibration \cite{spe}. The resulting PE number provides the energy estimator for the data. The PSD variable $F_{\rm prompt}$ distinguishes nuclear recoil events at high $F_{\rm prompt}$ from $^{39}$Ar decays and electron recoil backgrounds (ERB) at low $F_{\rm prompt}$. For this measurement it is defined as the fraction of PE detected in a time window of [-28, 150]\,ns around the event time and is calculated as
\begin{equation}
    F_{\rm prompt} = \frac{\sum_{t=-28\,{\rm ns}}^{150\,{\rm ns}} PE(t)}{\sum_{t=-28\,{\rm ns}}^{10\,{\rm \mu s}}PE(t) }.
\end{equation}
The ERB is composed of events generated by both $\gamma$-rays emitted by trace radioactivity in detector components and $\beta$-decays which scatter on electrons in the LAr. 

A more detailed description of the DEAP-3600 detector can be found in Ref. \cite{detpaper}.

\section{Liquid Argon Mass Estimate}\label{sec:mass}
The LAr mass is determined by evaluating both its density and its volume within the AV. This method previously resulted in a LAr mass of $(3279 \pm 96)$\,kg~\cite{1styearpaper}. That result has been refined for this work.

Two inputs are required to evaluate the volume of LAr in the detector: the AV radius and the LAr height within the AV. The internal radius of the AV was measured during its construction. After correcting for the thermal contraction that occurred during cool-down using a temperature-dependent coefficient measured in Ref. \cite{Hartwig1994} the AV radius is determined to be (845.6 $\pm$ 0.9)~mm.
The LAr height is measured by taking advantage of the total internal reflection of the UV light at the GAr/LAr interface. The TPB re-emits light isotropically and so the photon detection rates for each PMT depend on the area of visible TPB immersed in the LAr. The rates for every PMT ring are averaged and the distribution is fit with an analytic model of the corresponding immersed area. This method is validated by comparing the data to Monte Carlo simulations of $^{39}$Ar decays within the LAr while varying the simulated LAr height. The best fit is found at a LAr height of $(550 \pm 10)$~mm above the equator and is stable across the dataset. The systematic uncertainty on the LAr height is the dominant source of uncertainty for the LAr mass estimate. A cross-check using the position reconstruction of detected events to evaluate the LAr height provides a consistent result. In this cross-check, a template fit in the reconstructed vertical position of $^{39}$Ar decay events is performed by comparing simulations with different LAr height values to the data histogram.

The LAr density is a function of its temperature. This temperature is constrained by the liquid-vapor transition of the argon in the AV and by the liquid-vapor transition of the nitrogen in the cooling coil. As the pressure in both systems is constantly recorded, the average LAr temperature is known within a few degrees K, and thus the effective density can be established to 0.5\% precision.

The possibility of argon bubbles is also investigated, the presence of which would reduce the total mass of LAr. Using the behavior of nitrogen as a reference~\cite{Bubbles2010} a limiting case is considered where all of the exterior heat entering the LAr creates bubbles. This worst-case scenario indicates that at most 6.3~kg of LAr is displaced by bubbles.

A toy Monte Carlo sampling the probability distribution functions (PDFs) of the AV radius, the LAr height, the LAr density, and bubble displacement is used to determine the central value of the LAr mass and its uncertainty. Flat PDFs are used for the constraints on the LAr density and the bubble displacement, while the AV radius and LAr height PDFs are considered Gaussian. According to this method, during the data-taking period of this measurement the DEAP-3600 AV contained $m_{LAr} = (3269 \pm 24)$~kg of LAr.

\section{Data Selection and Livetime Calculation}\label{sec:datasel}

\subsection{Run Selection}\label{sec:runselection}
The dataset is divided into discrete runs during which signals from the LAr are recorded. A single run is typically about 22 hours long, though this can vary between just a few minutes and up to about 2 days. The runs examined here are from the same 2016-2017 dataset used for the dark matter search published by the DEAP collaboration~\cite{1styearpaper} with the additional restriction that runs are at least 18 hours long. This requirement is imposed to ensure sufficient statistics to fit the $\gamma$-dominated region of the ERB spectrum in each run.

The selection of runs is based on stability criteria concerning the cooling system of the AV, the charge distributions in the PMTs, and the efficiency of the trigger. A data cleaning cut is applied to each run to reject events occurring within $\delta t_{\rm cut} = 32\ \mu$s of the previous event, which removes $\delta t_{i} \leq 32 \mu$s of livetime per event $i$; the total number of events removed by this cut is $N_{\rm DCcut}$. Low-level cuts are then applied to reject events recorded from pulse injections by periodic monitoring triggers and events with inconsistent data acquisition readouts such as busy signals, for a total of $N_{\rm LLcut}$ events. The events removed by these cuts, along with all remaining physics triggers $N_{\rm phys}$, are taken into account in the run-dependent livetime calculation shown in Eq. \ref{eq:tlive}.
\begin{equation}
    \begin{aligned}
    T_{\rm live} = T_{\rm run} &- \sum_{i = 1}^{N_{\rm DCcut}} \delta t_{i} - N_{\rm LLcut} \cdot \delta t_{\rm cut} \\
    &- N_{\rm phys} \cdot (\delta t_{\rm cut} - \delta t_{\rm int}). 
    \end{aligned}
    \label{eq:tlive}
\end{equation}
Here, $T_{\rm live}$ is the livetime for a run, $T_{\rm run}$ is the total time of that run, and $\delta t_{\rm int} = 10\ \mu$s corresponds to the charge integration window during which the detector can record a pile-up event, while the time between $\delta t_{\rm int}$ and $\delta t_{\rm cut}$ is dead time.  The value of $N_{\rm phys}$ includes the prescaled triggers as the timestamp of each of these events is stored. Testing of the algorithm was performed using values of $\delta t_{\rm cut}$ ranging from 20 $\mu$s up to 250 $\mu$s for a selection of data runs.  For each $\delta t_{\rm cut}$ value the livetime and specific activity of each run were calculated. We observed negligibly small variations in the measured specific activity as a function of $\delta t_{\rm cut}$, as expected.

An offline reduction is applied where precisely 1 out of every 100 events from outside the prescaled trigger region is kept in order to remove boundary effects and obtain a smooth spectrum.

\subsection{Event Selection}\label{sec:eventselection}
In addition to the data cleaning and low-level cuts described in the previous section, event selection cuts are applied. Pile-up needs to be taken into account given the high rate of $^{39}$Ar decays and the length of the event window: approximately 5\% of recorded events are expected to contain 2 or more decays. Additionally, a triggered event can follow an energy deposit which occurred during time in which DAQ was busy and unable to record (deadtime). The late scintillation light from this previous, unrecorded energy deposit can reach into the beginning of the triggered event. Since the full energy of the previous energy deposit is not visible in the digitized trace, this type of pile-up is hard to model. While this analysis endeavours to keep pile-up events and account for them in the specific activity calculation, events with this pre-trigger pileup are not suitable analysis candidates.

To select events without pre-trigger pile-up the time at which the event occurred within the trigger window must be in the range [2250, 2700]~ns, and it is required that fewer than 4 pulses are recorded by the PMTs in the first 1600~ns of the event. These cuts do not remove a significant number of events, and the majority of the events removed are at very low energies. These removed events are mainly outside the range of the fits described in Section \ref{sec:energyfit}. 

Electron recoil events, which are dominated by $^{39}$Ar decays at lower energies and $\gamma$ backgrounds at higher energies, are selected with the requirement 0.1 $\leq$ $F_{\rm prompt}$ $\leq$ 0.5. These events, along with $^{39}$Ar-$^{39}$Ar pileup events and $^{39}$Ar signal events, are shown in Figure \ref{fig:ar39datafitsfreq}. A more in-depth discussion of the electron recoil events can be found in Ref. \cite{erb}.

This analysis also makes use of a peak-finding algorithm based which examines the PMT waveforms to count the number of ``sub-events'' within the trigger window. The algorithm counts pulses from each PMT in the event window to look for statistically significant increases in the pulse count and is able to identify sub-events separated by as little as 50 ns. When tested using MC the algorithm was able to correctly identify 96\% of $^{39}$Ar-$^{39}$Ar pile-up events, and only 0.1\% of single $^{39}$Ar were incorrectly identified as having multiple sub-events. The number of sub-events is used to select pile-up candidates in order to perform a data-driven estimate of the double $^{39}$Ar pile-up cut efficiency as described in Section~\ref{sec:calc}.

\section{Specific Activity Measurement}\label{sec:method}

The specific activity of $^{39}$Ar is measured individually for each run in the dataset. Each measurement is based on a fit to the low $F_{\rm prompt}$ energy spectrum and consists of an $^{39}$Ar $\beta$-decay spectrum (single $^{39}$Ar), a spectrum with two $^{39}$Ar decays occurring within the same trigger window (double $^{39}$Ar pile-up), and a spectrum containing all non-$^{39}$Ar ERB events scaled to the activities measured in Ref. \cite{erb}. 
The ERB and double $^{39}$Ar pile-up input spectra are generated by simulating events within the DEAP-3600 detector using the \texttt{RAT} software. The single $^{39}$Ar component is built directly from the theoretical model provided by Kostensalo et al. \cite{kostensalo}. 
Each of the three model components is normalized using a parameter in the fits. Energy scale PE and detector resolution effects $\sigma$(PE) in the form of a Gaussian term are applied to all three model components, parameterized as
\begin{equation}
    \begin{aligned}
    &PE = p_0 + p_1\cdot E + p_2\cdot E^2, \\
    &\sigma(PE) = \sqrt{p_3\cdot PE + p_4 \cdot PE^2}.
    \end{aligned}
\end{equation}
The constant energy scale parameter $p_0$ is fixed in the fits to a value obtained by measuring PMT baselines.  The number of $^{39}$Ar decays is split into two main components as
\begin{equation}
    N = N_{\rm single} + N_{\rm pile-up},
\end{equation}
\noindent with the number of single $^{39}$Ar decays $N_{\rm single}$ and the number of $^{39}$Ar decays which are part of a pile-up event $N_{\rm pile-up}$. The latter number includes double $^{39}$Ar decays, triple $^{39}$Ar decays (three $^{39}$Ar decays in one trigger window), pile-up of $^{39}$Ar decays with ERB decays, and pile-up of $^{39}$Ar decays with high $F_{\rm prompt}$ events ($F_{\rm prompt} > 0.5$).

\subsection{Fitting the Energy Spectrum}\label{sec:energyfit}

This analysis uses \texttt{Minuit} in \texttt{ROOT} \cite{minuit} to fit the three input spectra to data. The fit performs a chi-square minimization as
\begin{equation}
\chi^2 = \sum_i^{n_{b}} \left(\frac{M_i - D_i}{\sqrt{M_i}}\right)^2 + \mathcal{P},
\label{eq:chi2}
\end{equation}
with the number of bins $n_b$ in the data histogram, the data content $D_i$ in bin $i$ and the model contribution $M_i$. The parameter $\mathcal{P}$ is a penalty term applied to a shape nuisance parameter $a_0$, which corrects the theoretical $^{39}$Ar input spectrum linearly in energy to fit the data; it is constructed to account for the differences $\Delta a_0 = 0.01$ observed between the Kostensalo et al. \cite{kostensalo} and the Behrens \& Janecke \cite{behrensjanicke} $^{39}$Ar $\beta$-shapes and is calculated as
\begin{equation}
    \mathcal{P} = \frac{a_0}{\Delta a_0}.
\end{equation}
The closer $a_0$ is to zero, the more the spectrum fits the shape by Kostensalo et al. The fit model is adapted from the model used for the energy response fits in Ref. \cite{1styearpaper}.

\begin{figure}
\centering
\includegraphics[width=\columnwidth]{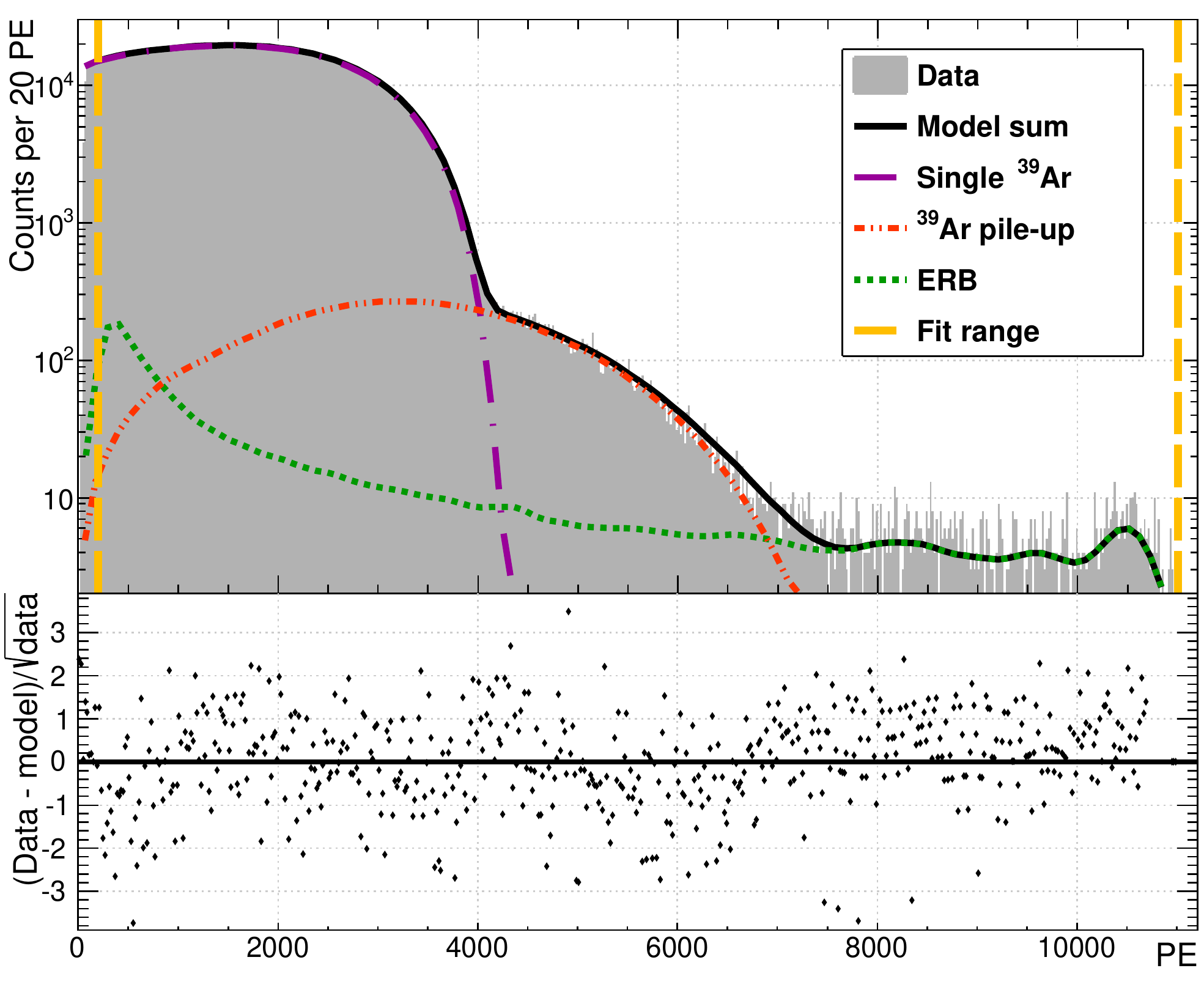}
 \caption{The top panel shows an example fit on one run including the $^{39}$Ar, ERB, and $^{39}$Ar pile-up components which form the fit function. The fit range from 200--11,000\,PE is shown by the vertical dashed lines.  The bottom panel shows the residual between the fit function and data normalized to the square root of the contents in the observed PE distribution. The fit is extrapolated below the lower bound to count events in the low-energy region where the trigger efficiency is not 100\%. The slight mismodelling of the ERB spectrum as apparent in the residual plot does not significantly affect the final result. The reduced chi-square for this run is given by $\chi^2/{\rm ndf} = 687.42/531 = 1.29$. The run shown here is approximately 28.5 hours long.}
 \label{fig:ar39datafitsfreq}
\end{figure}

The fit range for this measurement is [200, 11000]~PE and is chosen to avoid trigger efficiency effects at low PE and to provide a handle for the fit to scale $n_{\rm double}$ and $n_{\rm ERB}$ beyond the $^{39}$Ar endpoint at high PE. This range includes the $^{40}$K $\gamma$-emission peak at 1460\,keV \cite{erb} (approximately 10,500~PE) and allows the ERB spectrum normalization to be determined in the fit. This analysis is performed with $n_{b}$ corresponding to a bin width of $b = 20$~PE which was chosen to provide sufficient statistics to define the $^{40}$K peak. 
The $^{39}$Ar $\beta$-shape nuisance parameter is an output of the fit.  Details of the fit inputs and outputs are provided in Table~\ref{tab:fitinputfreq}, along with the other parameter values taken as input to the specific activity measurement. Figure \ref{fig:ar39datafitsfreq} shows an example fit using this model for one data run.
The parameters from each fit are examined to look for trends across the dataset and for issues such as getting stuck at the boundaries of their allowed ranges: no such issues are observed.

\begin{table*}
\centering
\caption{Parameters, their values and constraints, and the resulting contributions to the uncertainty for the specific activity measurement. 
Negligibly small systematic uncertainties are indicated with `--'. 
The dominant uncertainty on $S_{\rm Ar39}$ arises from the uncertainties on event selection cut efficiency values as determined with the data-driven method (d-d) and the Monte Carlo method (MC).
}
\label{tab:fitinputfreq}
\begin{tabular*}{\textwidth}{{@{\extracolsep{\fill}}lllll@{}}}
\hline\noalign{\smallskip}
\textbf{Parameter} & \textbf{Symbol} & \textbf{Value} & \textbf{Constraints} & \textbf{Absolute uncertainty}  \\  
 & & & & \textbf{on ${\bf S_{\bf Ar39}}$ [Bq/kg${\bf _{\bf atmAr}}$]} \\
\noalign{\smallskip}\hline\noalign{\smallskip}
Fit range &  & [200, 11000]~PE & Fixed & 0.001 \\ 
Histogram bin width & $b$ & 20\,PE & Fixed & 0.001 \\
\noalign{\smallskip}\hline\noalign{\smallskip}
Constant energy scale parameter & $p_0$ & $(1.3 \pm 0.4)$ PE & Fixed & -- \\ 
Linear energy scale term & $p_1$ & [7.1, 7.3] PE/keV & Free-floating, run-dependent & 0.009 \\
Quadratic energy scale term & $p_2$ & -- & Not considered in this method & -- \\
Linear resolution parameter & $p_3$ & [1.67, 1.73] PE & Free-floating, run-dependent & 0.009 \\ 
Quadratic resolution parameter & $p_4$ & [2.1, 3.8] $\times 10^{-4}$ & Free-floating, run-dependent & 0.001 \\
\noalign{\smallskip}\hline\noalign{\smallskip}
$^{39}$Ar $\beta$-shape nuisance parameter & $a_0$ &  & Free-floating,  & 0.001 \\
& & & ~~~~~~constrained by a penalty term & \\
$^{39}$Ar normalization & $n$ &  & Free floating, run-dependent & -- \\ 
Double $^{39}$Ar pile-up normalization & $n_{\rm double}$ &  & Free floating, run-dependent & -- \\
ERB normalization & $n_{\rm ERB}$ & & Free-floating, run-dependent & -- \\
$^{85}$Kr normalization & $n_{{\rm Kr85}}$ & & Upper limit, see Section~\ref{sec:calc}  & 0.010 \\
\noalign{\smallskip}\hline\noalign{\smallskip}
Liquid argon mass & $m_{LAr}$ & $(3269 \pm 24)$~kg & Measured, see Section~\ref{sec:mass} & 0.007 \\
Live-time & $T_{\rm live}$ & 167 d [sum of all runs] & Measured, see Section~\ref{sec:datasel} & -- \\
 \noalign{\smallskip}\hline\noalign{\smallskip}
Cut efficiency on single $^{39}$Ar & $\epsilon$ & 0.983 [d-d], 0.999 [MC] & \multirow{2}{*}{Measured, see Section~\ref{sec:calc}} & \multirow{2}{*}{0.016} \\
Cut efficiency on double $^{39}$Ar pile-up & $\epsilon_{\rm double}$ & Run- \& energy-dependent &  &  \\
\noalign{\smallskip}\hline
\end{tabular*}
\end{table*}

%------------------------------------------------------------------------------------------------------------------------------------

\subsection{Calculating the Specific Activity}\label{sec:calc}

The number of $^{39}$Ar single decays $N_{\rm single}$ is obtained from the single $^{39}$Ar spectrum fit result integral $n$ as 
\begin{equation}
    N_{\rm single} = \frac{n \cdot a_{\rm presc}}{\epsilon \cdot b},
    \label{eq:singlecomp}
\end{equation}
with the bin width $b$ of the fitted data histogram, a trigger prescaling correction factor $a_{\rm presc} = 100$ and the cut efficiency $\epsilon$. The main, data-driven method used to estimate this cut efficiency involves defining a loose event selection for the denominator spectrum of events present before the cuts with an $F_{\rm prompt} < 0.7$ requirement; this loose event selection removes the unwanted Cherenkov and nuclear recoil events. The numerator spectrum for the efficiency calculation contains those events which pass the selection cuts described in Section~\ref{sec:eventselection}.  

First, the efficiency $\epsilon_{\rm lowerPE}$ is calculated over the range [300, 3000]~PE, which is dominated by single $^{39}$Ar events, by taking the ratio of the two spectra bin-by-bin. Then, to extract the efficiency $\epsilon$ for single $^{39}$Ar events, a correction is applied to $\epsilon_{\rm lowerPE}$ to account for the presence of double $^{39}$Ar pile-up in the sample.
The cut efficiency $\epsilon_{\rm double}$ is calculated bin-by-bin in the PE histogram of each run over the range [300, 3000]~PE with an additional event selection requirement for the numerator and denominator spectra to select events which contain exactly two sub-events.
The efficiency for single $^{39}$Ar events is calculated bin-by-bin by next solving for $\epsilon_i$ using the bin-dependent $\epsilon_{\rm lowerPE}$ and $\epsilon_{\rm double}$ values in the following equation as
\begin{equation}
    \epsilon_{{\rm lowerPE}, i} = \epsilon_i\cdot x_i + \epsilon_{{\rm double}, i}\cdot (1 - x_i),
    \label{eq:epsilon}
\end{equation}
where $x_i$ is the fraction of single $^{39}$Ar events measured by the fits in bin $i$. The value for $\epsilon$ is then calculated as the average of the $\epsilon_i$ values as any energy dependence here is negligible. The resulting data-driven value of $\epsilon$ is calculated run-by-run and each value is used in the calculations for its respective run. The average $\epsilon$ is 0.983.

As a cross-check to this data-driven method, the cut efficiency values are evaluated using the Monte Carlo simulated samples of the $^{39}$Ar decays and the double $^{39}$Ar pile-up described earlier.  While the simulations do not describe the data perfectly, this method yields clean spectra of these two event classes which can be individually analyzed.  With the Monte Carlo method $\epsilon_{\rm MC} = 0.999$. We evaluate the specific activity $S_{\rm Ar39}$ using both $\epsilon$ and $\epsilon_{\rm MC}$ as inputs to Eq.~\ref{eq:singlecomp} and take the difference as a systematic uncertainty.  This difference is the dominant source of systematic uncertainty for this measurement.

The number of $^{39}$Ar decays that are part of pile-up events is split into the different components as
\begin{equation}
    \begin{aligned}
    N_{\rm pile-up} &= N_{\rm double} + N_{\rm triple} + N_{\rm ERB,Ar39} + N_{\rm hFp,Ar39}.
    \end{aligned}
    \label{eq:pileupcomp}
\end{equation}
Here, $N_{\rm double}$ is the number of $^{39}$Ar decays that are part of a double $^{39}$Ar pile-up event, $N_{\rm triple}$ is the number of $^{39}$Ar decays that are part of a triple pile-up event, $N_{\rm ERB,Ar39}$ is the number of $^{39}$Ar decays which pile-up with a ERB recoil, and $N_{\rm hFp,Ar39}$ is the number of $^{39}$Ar decays which pile-up with a high $F_{\rm prompt}$ process such as Cherenkov light or a nuclear recoil. $N_{\rm double}$ is obtained from the double $^{39}$Ar spectrum fit result integral $n_{\rm double}$ as
\begin{equation}
    N_{\rm double} = \frac{n_{\rm double} \cdot a_{\rm presc}}{\epsilon_{\rm double} \cdot b} \cdot 2.
    \label{eq:doublecomp}
\end{equation}
Here, $\epsilon_{\rm double}$ is the cut efficiency on double $^{39}$Ar pile-up events described above, and the factor 2 corrects for 2 $^{39}$Ar decays in 1 double pile-up event.  
The energy-dependence of $\epsilon_{\rm double}$ over the wider range of the pile-up spectrum is taken into account by applying this correction bin-by-bin.

$N_{\rm double}$ is utilized to calculate the single $^{39}$Ar rate $R_{\rm Ar39}$ from a first-order pile-up calculation as shown in Eq. \ref{eq:arrate}.
\begin{equation}
    R_{\rm Ar39} = \sqrt{\frac{N_{\rm double}}{2 \cdot T_{\rm live} \cdot \delta t_{\rm int}}}.
    \label{eq:arrate}
\end{equation}
$R_{\rm Ar39}$ is used to determine the remaining pile-up components which are estimated with first-order pile-up approximations as
\begin{equation}
    \begin{aligned}
        N_{\rm triple} &= 3\cdot R_{\rm Ar39}^3 \cdot \delta t_{\rm int}^2 \cdot T_{\rm live}, \\
        N_{\rm ERB,Ar39} &= R_{\rm Ar39} \cdot R_{\rm ERB} \cdot \delta t_{\rm int} \cdot T_{\rm live}, \\ 
        N_{\rm hFp,Ar39} &= R_{\rm Ar39} \cdot R_{\rm hFp} \cdot \delta t_{\rm int} \cdot T_{\rm live},
    \end{aligned}
    \label{eq:rempileup}
\end{equation}
where the factor of 3 in $N_{\rm triple}$ accounts for the 3 $^{39}$Ar decays in each of these pile-up events. The ERB rate $R_{\rm ERB} = (10.5 \pm 0.6)$\,Hz and the high $F_{\rm prompt}$ rate $R_{\rm hFp} = (270 \pm 3)$\,Hz are established from the fit output $n_{\rm ERB}$ and from the rate of events observed in the high $F_{\rm prompt}$ window in the dataset, respectively.  The pile-up rates can be calculated by dividing the quantities in Eq. \ref{eq:rempileup} by $T_{\rm live}$.

Beyond the ERB measured in Ref. \cite{erb}, the dataset considered in this analysis may contain a small number of $^{85}$Kr $\beta$-decay events. The $^{85}$Kr beta spectrum has an endpoint energy of 687.0~keV; this is in the region dominated by the double $^{39}$Ar pileup events.
Uncertainty in the amplitude of a peak at 600.66~keV from the $^{226}$Ra chain makes obtaining the $^{85}$Kr from fitting the energy spectrum challenging.
The $^{85}$Kr activity is studied a posteriori by repeating the fit including the $^{85}$Kr $\beta$-shape from Ref. \cite{krshape} with a normalization parameter $n_{\rm Kr85}$, while also varying the energy response parameters and the $^{39}$Ar endpoint within their uncertainties. No cuts are made to remove the double $^{39}$Ar pileup so that both the nominal fits and the fits including a $^{85}$Kr spectrum are performed on the same data.
These fit results suggest that at most 0.01~Bq/${\rm kg_{atmAr}}$ of $^{85}$Kr is present in our dataset.  This limit is considered as an additional source of systematic uncertainty.

\subsection{Results}
\label{sec:results}

\begin{figure}
\centering
\includegraphics[width=\columnwidth]{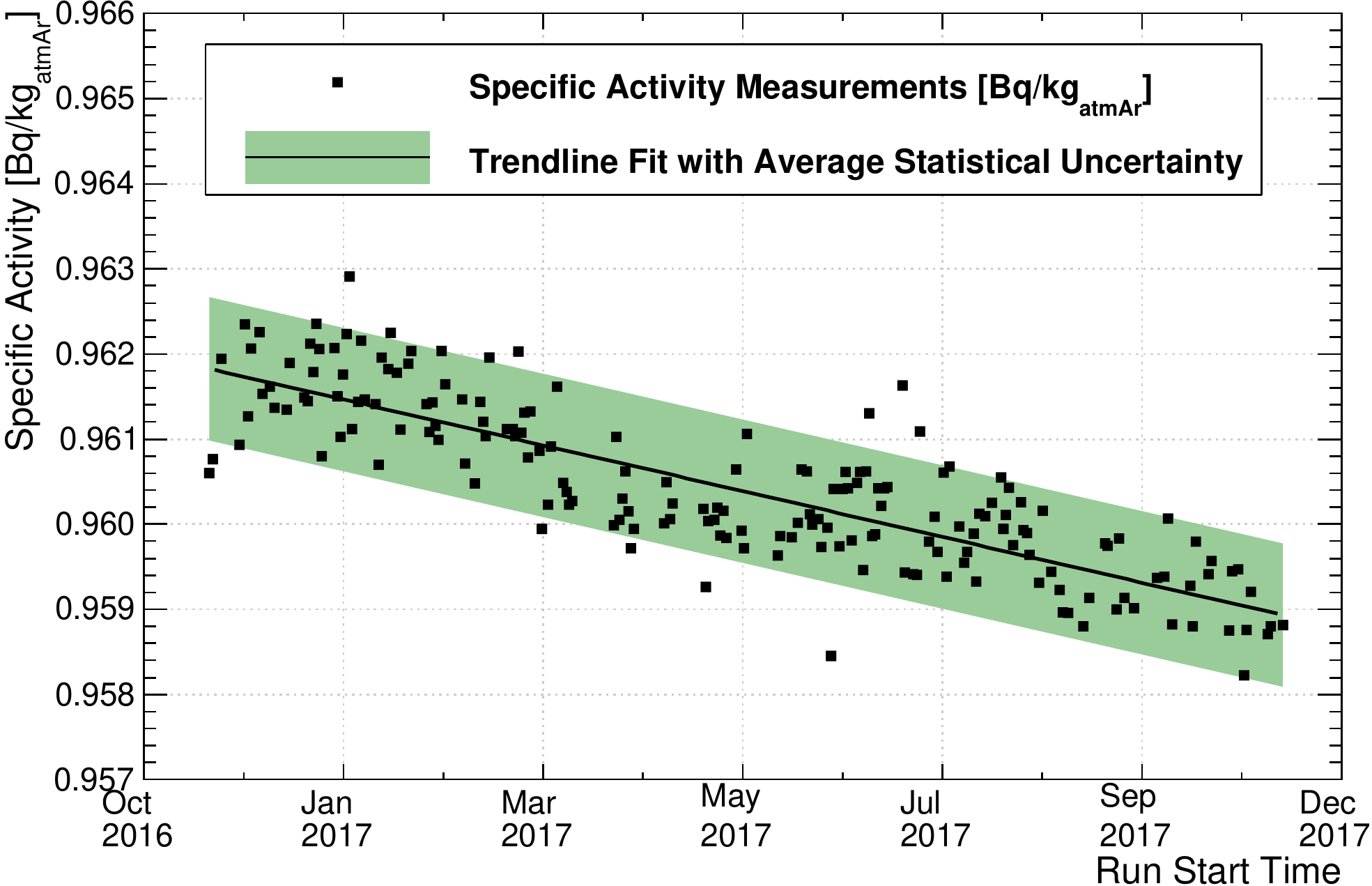}
 \caption{The measured specific activity of $^{39}$Ar versus run time for the entire dataset. The exponential trendline fit is shown, with the average statistical uncertainty depicted as an error band. The systematic uncertainty band is wider than the y-axis range shown here.
 }
 \label{fig:specactresultsfreq}
\end{figure}

The specific activity of $^{39}$Ar is evaluated for each run by combining Eqs.~\ref{eq:singlecomp}, \ref{eq:doublecomp} and \ref{eq:rempileup} with Eq.~\ref{eq:specact}.
The run-by-run results are presented in Figure \ref{fig:specactresultsfreq} which includes an exponential fit to the measured specific activity over time.  This fit is used to determine the specific activity value at the start of the dataset.

Uncertainties due to the liquid argon mass estimate and related to the determination of cut efficiencies were discussed in Sections~\ref{sec:mass} and~\ref{sec:calc} respectively.
Additional systematic uncertainties on the specific activity measurement are evaluated as follows.
For each run, the fit is repeated with the linear energy scale parameter $p_1$ fixed to its central value for that run plus or minus 0.15 PE/keV. 
The uncertainties on the other energy scale and resolution parameters $p_0$, $p_3$ and $p_4$ are likewise propagated to the measurement by repeating the fits with fixed parameters set according to their uncertainties determined in the energy response measurement described in Ref.~\cite{1styearpaper}.
Uncertainties due to the choice of the histogram bin width (varied to 10~PE and to 40~PE) and the choice of the fit range (lower bound increased to 500 PE) are evaluated in a similar manner. Theoretical $\beta$-shape uncertainties are accounted for by repeating the fit with $a_0$ fixed to 0, and then fixed to the median value found over the entire dataset. The systematic uncertainty due to the ERB normalization is negligible, and so any systematics associated with the MC generation of the spectra used in the fits is similarly negligible. Optical model uncertainties within the MC do not affect the pile-up spectrum shape used in the fits. The systematic uncertainty due to the double $^{39}$Ar pile-up spectrum shape and normalization are negligible.
The impact of each source of systematic uncertainty on the result is detailed in Table~\ref{tab:fitinputfreq}.

The statistical uncertainty of 0.001~Bq/kg$_{\rm atmAr}$ shown in Figure \ref{fig:specactresultsfreq} is calculated by propagating the uncertainties on the ERB and the high $F_{\rm prompt}$ background rates, and the fit uncertainties on the single $^{39}$Ar and the double $^{39}$Ar pile-up normalization parameters. The fit uncertainty of the $^{39}$Ar spectrum dominates the statistical uncertainty.

A correction is applied to the measured specific activity determined from the exponential fit to account for the age of the argon.  The correction factor is calculated as
\begin{equation}
    \eta_t = 2\hat{\mkern6mu}({t_{\rm age}}/{T_{1/2}}),
    \label{eq:timeCorr}
\end{equation}
where $t_{\rm age} = (1.0 \pm 0.5$)~y is the average time between atmospheric extraction of the argon and the start of the data-taking period. Multiplying by $\eta_t$ corrects for the approximately 0.26\,\% drop in the activity before data were taken.
Cosmogenic activation of $^{39}$Ar during the time after the argon was extracted from the atmosphere is negligible.

The specific activity of $^{39}$Ar in atmospheric argon is measured to be
\begin{equation*}
     S_{\rm Ar39} = (0.964 \pm 0.001_{\rm stat} \pm 0.024 _{\rm sys}) \,\textrm{Bq/kg}_{\rm atmAr}.
\end{equation*}

\vspace{-3pt}
\section{Cross-Check Analysis}\label{sec:crosscheck}

Here we present a cross-check to our result which is an update of an earlier analysis, the details of which are described in Ref.~\cite{mattthesis}. This analysis used the Bayesian Analysis Toolkit (BAT) \cite{bat} software to fit the input spectra to the data and extract the model parameters. BAT uses Markov Chain Monte Carlo to generate posterior probability distributions of the fit parameters based on prior probability distributions and a likelihood function input by the user. This cross-check also differs from the analysis presented in previous sections by applying a different set of event selection cuts than those described in Section \ref{sec:eventselection}. A cut on the event time within the trigger window was not applied, and the peak-finding algorithm to count ``sub-events'' was used to remove the majority of pileup events. The data cleaning cut to remove events close in time to a previous event was not applied.
Otherwise, the same criteria described in Section \ref{sec:datasel} were applied and a fit was performed on each run in the dataset.

The detector response model in the fit included a constant energy scale parameter $p_0$, a linear energy scale parameter $p_1$, a quadratic energy scale parameter $p_2$, and a linear resolution parameter $p_3$.  The quadratic resolution parameter $p_4$ was not considered in these fits.  The $p_1$, $p_2$, and $p_3$ model parameters were given flat priors in the fits and allowed to float. The $p_0$ parameter was fixed in the fits.

The nominal input $^{39}$Ar spectrum used was from Behrens \& Janecke \cite{behrensjanicke}. Each fit returned the normalization of this spectrum and was given a flat prior. In addition to the $^{39}$Ar spectrum, the inputs to the fit were an ERB spectrum and an MC-generated spectrum of double $^{39}$Ar pileup events which survive the cuts. Each fit returned the normalization parameters for these spectra; at the input stage these were given Gaussian priors with a mean value of 1, which corresponded to a normalization based on an assumed event rate and the known runtime. All three normalization parameters were allowed to float in the fits, and the posterior values were used to calculate the $^{39}$Ar specific activity. The single $^{39}$Ar events counted through the fit outputs of this cross-check method and that described in Section \ref{sec:energyfit} do not differ significantly. An additional set of fits was performed using the Kostensalo et al. spectrum \cite{kostensalo}.  The measured specific activity differed by a negligible amount between these fits and those using the nominal spectrum.

The result in Ref. \cite{mattthesis} has been updated here to include the updated LAr mass, the new data-driven cut efficiency estimates, the revised livetime calculation, and the correction for the age of the argon. This method yields the following value for the specific activity of $^{39}$Ar at the time of atmospheric extraction: 
$(0.97 \pm 0.001_{\rm stat} \pm 0.03_{\rm sys}) \,\textrm{Bq/kg}_{\rm atmAr}$.

%------------------------------------------------------------------------------------------------------------------------------------

\begin{table}
\centering
\caption{Summary of specific activity measurements of $^{39}$Ar by different collaborations. }
\label{tab:resultsummary}
\resizebox{\columnwidth}{!}{
\begin{tabular}{ll}
\hline\noalign{\smallskip}
\textbf{Measurement} & \textbf{Specific activity [Bq/kg$_{\bf atmAr}$]} \\ 
\noalign{\smallskip}\hline\noalign{\smallskip}
WARP \cite{warp} & 1.01 $\pm$ 0.02$_{\rm stat}$ $\pm$ 0.08$_{\rm sys}$ \\
ArDM \cite{ardm} & 0.95 $\pm$ 0.05 \\
DEAP-3600 (this work) & 0.964 $\pm$ 0.001$_{\rm stat}$ $\pm$ 0.024$_{\rm sys}$ \\
\noalign{\smallskip}\hline
\end{tabular}
}
\end{table}

\section{Conclusion}\label{sec:conclusion}

A measurement of the specific activity of $^{39}$Ar in atmospheric argon using the LAr target of the DEAP-3600 detector has been presented. This result is the most precise measurement of the specific activity of $^{39}$Ar in atmospheric argon to date and agrees with existing measurements which are summarized in Table \ref{tab:resultsummary}. 
The high precision of this measurement is owing to a combination of factors including the low-background nature of DEAP-3600, the large number of decays observed in each data run, and the precise measurement of the LAr target mass. 
The statistical uncertainties here are much smaller than the systematic uncertainties due to the high statistics of the data. The dominant systematic uncertainties arise from the event selection cut efficiencies, the energy scale and energy resolution parameters, and the possible presence of $^{85}$Kr within the LAr. 

This precision measurement is an important input to the background models of experiments operating with argon as a medium. It will benefit current experiments, help to inform the design of future detectors, and support measurements in radiometric dating which use the $^{39}$Ar/Ar ratio as an input.

\begin{acknowledgements}

We thank the Natural Sciences and Engineering Research Council of Canada (NSERC),
the Canada Foundation for Innovation (CFI),
the Ontario Ministry of Research and Innovation (MRI), 
and Alberta Advanced Education and Technology (ASRIP),
the University of Alberta,
Carleton University, 
Queen's University,
the Canada First Research Excellence Fund through the Arthur B.~McDonald Canadian Astroparticle Physics Research Institute,
%Consejo Nacional de Ciencia y Tecnología Project No. CONACYT CB-2017-2018/A1-S-8960, 
%DGAPA UNAM Grants No. PAPIIT IN108020 and IN105923,
%and Fundación Marcos Moshinsky.
Consejo Nacional de Ciencia y Tecnolog\'ia Project No. CONACYT CB-2017-2018/A1-S-8960, 
DGAPA UNAM Grants No. PAPIIT IN108020 and IN105923, 
and Fundaci\'on Marcos Moshinsky,
the European Research Council Project (ERC StG 279980),
the UK Science and Technology Facilities Council (STFC) (ST/K002570/1 and ST/R002908/1),
the Leverhulme Trust (ECF-20130496),
the Russian Science Foundation (Grant No. 21-72-10065),
the Spanish Ministry of Science and Innovation (PID2019-109374GB-I00) and the Community of Madrid (2018-T2/ TIC-10494), 
the International Research Agenda Programme AstroCeNT (MAB/2018/7)
funded by the Foundation for Polish Science (FNP) from the European Regional Development Fund,
and the European Union's Horizon 2020 research and innovation program under grant agreement No 952480 (DarkWave).
Studentship support from
the Rutherford Appleton Laboratory Particle Physics Division,
STFC and SEPNet PhD is acknowledged.
We thank SNOLAB and its staff for support through underground space, logistical, and technical services.
SNOLAB operations are supported by the CFI
and Province of Ontario MRI,
with underground access provided by Vale at the Creighton mine site.
We thank Vale for their continuing support, including the work of shipping the acrylic vessel underground.
We gratefully acknowledge the support of the Digital Research Alliance of Canada,
Calcul Qu\'ebec,
the Centre for Advanced Computing at Queen's University,
and the Computational Centre for Particle and Astrophysics (C2PAP) at the Leibniz Supercomputer Centre (LRZ)
for providing the computing resources required to undertake this work.

\end{acknowledgements}

\bibliographystyle{spphys}
\bibliography{bibliography}

\end{document}